\documentclass[12pt]{iopart}
\usepackage{iopams} 
\usepackage{bm}
\usepackage{graphicx}
\usepackage{hyperref}
\hypersetup{
	colorlinks=true,
	linkcolor=blue,
	urlcolor=blue,
	citecolor=blue,
}

\eqnobysec 

\begin{document}
\title{Single-ion addressing via trap potential modulation in global optical fields}
\author{Christopher M. Seck$^*$\footnote{Present address: Computational Sciences and Engineering Division,
		Oak Ridge National Laboratory, Oak Ridge, TN 37831, USA}, Adam M. Meier, J. True Merrill, Harley T. Hayden, Brian C. Sawyer, Curtis E. Volin, Kenton R. Brown$^\dagger$}
\address{Georgia Tech Research Institute, Atlanta, Georgia 30332, USA}
\eads{\mailto{$^*$\href{mailto:seckcm@ornl.gov}{seckcm@ornl.gov}}, \mailto{$^\dagger$\href{mailto:kenton.brown@gtri.gatech.edu}{kenton.brown@gtri.gatech.edu}}}
\vspace{10pt}
\begin{indented}
	\item[]22 January 2020
\end{indented}

\begin{abstract}
To date, individual addressing of ion qubits has relied primarily on local Rabi or transition frequency differences between ions created via electromagnetic field spatial gradients or via ion transport operations. Alternatively, it is possible to synthesize arbitrary local one-qubit gates by leveraging local phase differences in a global driving field. Here we report individual addressing of $^{40}$Ca$^+$ ions in a two-ion crystal using axial potential modulation in a global gate laser field. We characterize the resulting gate performance via one-qubit randomized benchmarking, applying different random sequences to each co-trapped ion. We identify the primary error sources and compare the results with single-ion experiments to better understand our experimental limitations. These experiments form a foundation for the universal control of two ions, confined in the same potential well, with a single gate laser beam.
\end{abstract}


\submitto{\NJP}

\maketitle


\section{Introduction}
The ability to produce arbitrary one-qubit rotations on each individual qubit is a requirement for a universal quantum computer \cite{Khaneja2001}. Single-qubit addressing is also crucial to characterization of quantum processes via quantum process tomography (QPT), randomized benchmarking (RB), and related techniques. Regardless of the physical qubit implementation, single-qubit addressing requires either (1) a differential Rabi frequency, (2) a differential qubit frequency, or (3) a phase shift at each of the qubit sites.

Ion-trap and neutral-atom optical-lattice systems have achieved differential Rabi frequencies using tightly focused laser beams where the beam waists are smaller than the typical interatom spacings. However, closely spaced atoms (separations are typically 1-4 $\mu$m in ion-trap systems) are required for the fastest and highest fidelity entangling gates. Maintaining a high degree of optical isolation between such tightly spaced qubit locations is challenging. In practice, the atoms often are moved to larger separations prior to one-qubit gates \cite{Kielpinski2002,Herold2016}, or a composite pulse sequence is used to compensate the effect of finite beam waist on neighboring qubits \cite{Ivanov2011, Merrill2014}. The first option isolates neighboring qubits from neighboring laser beams, but the required ion transport operations can be costly both in time and in atom motional excitation \cite{Kaufmann2014}. The second option increases operation time, can increase error on the target qubit, and still requires tightly focused beams.

An alternative addressing technique in trapped-ion systems relies on micromotion-dependent Rabi frequencies. Here, an ion is displaced within the trap's radio-frequency electric-field gradient to alter the Rabi frequency of a micromotion sideband \cite{Leibfried1999}, and this sideband transition is driven with a global laser beam. However, experimental demonstrations report errors of ${\approx}10^{-2}$, the technique suffers from pronounced sensitivity to the local electrostatic environment (requiring frequent recalibration), and the requirement of displacing only one ion within a longer chain places undesired constraints on the trapping potentials \cite{Navon2014,Akerman2015}.

Differential qubit-frequency shifts have also been achieved with an auxiliary field gradient that generates spatially dependent Zeeman or Stark shifts. Here, individual addressing is accomplished by tuning the control field to the local qubit resonance; related techniques have been demonstrated in neutral atom \cite{Weitenberg2011}, ion-trap \cite{Warring2013,Johanning2009,Wang2009}, solid-state, and superconducting qubit \cite{Majer2007,Sillanpaa2007,Blais2004} experiments. Accurate spectral resolution of individual qubits imposes fundamental limits on the minimum gate time with these schemes because of the finite Fourier frequency width of a gate pulse. Additionally, the number of qubits that can be individually addressed is limited by the frequency tuning range of the control field, and the auxiliary field adds complexity and requires a precisely controlled amplitude. Furthermore, these gradient techniques are incompatible with long-coherence, field-insensitive qubit transitions \cite{Langer2005}.

Tightly-focused optical beams can also be used to impart a differential phase at individual qubit sites \cite{Schindler2013}, but such an approach faces similar challenges to those using differential Rabi frequencies. Alternatively, differential phase shifts can be achieved in a global beam through changes in ion position. Basic demonstrations of this idea have been performed both with two widely spaced ions \cite{Gutierrez2017} as well as with ions confined within the same potential minimum \cite{Kielpinski2001,Rowe2001,Schaetz2004}.

Here we generalize these concepts to achieve arbitrary, individually addressed one-qubit rotations on a two-ion crystal, leveraging controlled changes in trapping confinement to vary the spacing between ions. This differential phase technique does not require tightly focused beams, ion chain split and merge operations, global inhomogeneous auxiliary fields, or sensitive ion micromotion sidebands, and it may be readily incorporated into other ion trapping experiments using common laboratory equipment. The ability to perform individual qubit addressing with existing equipment will greatly increase the accessibility and approachability of multi-qubit gate characterization techniques, such as RB and QPT, in trapped-ion experiments that currently rely on more limited Bell-state fidelity analysis.

In this manuscript we describe the theory of single-ion addressing via trap potential modulation in global optical fields (\sref{sec:Theory}), and we detail our experimental apparatus (\sref{sec:Apparatus}) including the GTRI-fabricated ion trap used to validate the theory. We present the results of individually-addressed Ramsey experiments on a pair of ions in the same potential minimum (\sref{sec:Ramsey}), and we further characterize the performance of the technique using RB with different random gate sequences applied to each ion (\sref{sec:2IonRB}). As an additional diagnostic, we compare the results of these experiments to the results of experiments on a single ion (\sref{sec:ErrorDiagnostics}).

\section{Theory of Operation}\label{sec:Theory}
Differential phase shifts can be generated by varying the positions of trapped-ion qubits between gate pulses. Laser-cooled ions naturally form ordered crystals, and in this crystalline phase, ions are trapped near equilibrium positions controlled by the trap geometry and operating parameters \cite{Yan2016}. For the special case of two identical ions forming a linear crystal in a harmonic Paul trap, the equilibrium ion separation is
\begin{equation}
d_0 = [Z^2 e_0^2/(2 \pi \epsilon_0 m \omega_0^2)]^{1/3}
\end{equation}
where $e_0$ is the elementary charge, $\epsilon_0$ is the permittivity of free space, $m$ and $Z$ are the one-ion mass and charge, and $\omega_0/2\pi$ is the axial secular frequency of a single ion in the harmonic well. If the trap secular frequency is changed by a value of $\Delta\omega$, the ion separation becomes
\begin{equation}
d = d_0 \left(1-\left[\frac{1}{\frac{\Delta \omega}{\omega_0} + 1}\right]^{2/3}\right).
\end{equation}
Given $k_z$ is the projection of the laser wavevector along the crystal axis, making such a secular frequency change between laser pulses leads to an effective differential phase shift of $\Delta\phi = k_z (d-d_0)$ between the two ions for a second pulse (with respect to the phases experienced in a pulse before the secular frequency change). As a concrete example, consider two $^{40}$Ca$^+$ ions confined in a 2 MHz harmonic potential with a 729 nm quadrupole-transition gate beam oriented at $45^{\circ}$ to the axis. A shift of $\Delta\omega/\omega_0 = \left\{^{+0.27}_{-0.19}\right.$ here is sufficient to produce a differential $\pi$ phase shift and is sufficiently small that the ion pair can remain stably trapped.

Such a differential phase can be used to construct arbitrary one-qubit rotations (achieving different rotations on each ion in the pair) even with a global laser beam. A straightforward construction of this kind consists of an initial laser pulse, a controlled change in ion separation, and a final laser pulse as follows: the first pulse achieves a rotation
\begin{equation}
U_1 = R^{(1)}(\theta_1,\phi)R^{(2)}(\theta_2,\phi)
\end{equation}
where
\begin{equation}
R^{(k)}(\theta,\phi) = exp[-i\theta(X_k\cos(\phi)+Y_k\sin(\phi))/2]
\end{equation}
is the Bloch sphere rotation operator and $X_k$ and $Y_k$ are the Pauli operators for qubit $k$. The trap potential is then adjusted to produce a differential phase shift $\Delta\phi = \pi$, so that the second laser pulse effects a modified rotation
\begin{equation}
U_2= R^{(1)}(\theta_1^{\prime},\phi^{\prime})R^{(2)}(\theta_2^{\prime},\phi^{\prime}+\pi),
\end{equation}
where $\phi^{\prime}$ can be fixed to $\phi$ by changing the global laser phase. The net unitary is then
\begin{equation}
U = U_2 U_1 = R^{(1)}(\theta_1+\theta_1^{\prime},\phi)R^{(2)}(\theta_2-\theta_2^{\prime},\phi).
\end{equation}
If the second laser pulse duration is chosen such that $\theta_2 = \theta_2^{\prime}$ (including compensation for possible spatial gradients in the laser beam intensity), the resulting net gate rotation is
\begin{equation}
U = U_2 U_1 = R^{(1)}(\theta_1+\theta_1^{\prime},\phi) I^{(2)}
\end{equation}
where $I^{(2)}$ is the identity gate on the second ion. By choosing $\phi$ and scaling $\theta_1$ and $\theta_1^{\prime}$ appropriately, an arbitrary rotation on the Bloch sphere can ideally be realized on the first ion without affecting the second ion. A different choice of $\phi^{\prime} = \phi + \pi$ allows for a rotation on only the second ion.

In the absence of background electric fields, the center-of-mass (COM) motional mode is not displaced by an overall change in axial secular frequency (realized by uniformly scaling the DC trap potentials). However, changes in confinement do displace the out-of-phase breathing mode, will displace the COM mode in the presence of stray electric fields, and can produce motional squeezing of both modes \cite{Heinzen1990,Wineland1994}. If these displacements and squeezing were sufficiently large, the resulting motional excitation would degrade subsequent coherent laser operations. The excitation from squeezing should be negligible: 0.02 phonons would result from an immediate potential change of $\Delta\omega/\omega_0 = 0.25$ on a ground-state-cooled mode. Displacement excitations may not be negligible and must be considered more carefully. Center-of-mass displacements can be eliminated through compensation of stray electric fields near the ions. Breathing-mode displacements can be suppressed by performing confinement potential changes adiabatically (more slowly than a mode period), either through intentional waveform design or by relying on the finite bandwidth of the ion trap electrode filters. Any remaining displacement can be removed via an impulsive “kick” to the ions with a duration shorter than a mode period. With trap frequencies of a few MHz, this implies that individual ion addressing via confinement potential modulation can be performed in a duration of a few $\mu$s, commensurate with the duration of the individual laser pulses involved. In practice, we find ourselves limited by our ability to cancel stray electric fields which lead to excitation of the COM mode (discussed in detail below), and we observe only negligible excitation of the breathing mode. 

While our primary focus is on the universal control of two ions, the above concepts can be extended to more than two ions co-trapped in a linear chain. In a three-ion chain, a $\pi$ phase shift is realized between adjacent (edge) ions with an adjustment $\Delta\omega/\omega_0 = \left\{^{+0.32}_{-0.21}\right.\left(^{+0.14}_{-0.12}\right)$ in the axial potential, only slightly larger than the adjustment for two ions. Here we outline one possible sequence, assuming equal beam intensities at each ion for simplicity and choosing four pulses explicitly:
\begin{equation}
U_1 = R^{(1)}(\theta_1,\phi_1)R^{(2)}(\theta_1,\phi_1)R^{(3)}(\theta_1,\phi_1)
\end{equation}
\begin{equation}
U_2 = R^{(1)}(\theta_1,\phi_1)R^{(2)}(\theta_1,\phi_1+\frac{\pi}{2})R^{(3)}(\theta_1,\phi_1+\pi)
\end{equation}
\begin{equation}
U_3 = R^{(1)}(\theta_2,\phi_2)R^{(2)}(\theta_2,\phi_2)R^{(3)}(\theta_2,\phi_2)
\end{equation}
\begin{equation}
U_4 = R^{(1)}(\theta_2,\phi_2-\pi)R^{(2)}(\theta_2,\phi_2)R^{(3)}(\theta_2,\phi_2+\pi).
\end{equation}
The net unitary is
\begin{equation}
U = U_{4}U_{3}U_{2}U_{1}
\end{equation}
\begin{equation}	
U = U_{4}U_{3}R^{(1)}(2\theta_1,\phi_1)R^{(2)}(\theta_1,\phi_1+\frac{\pi}{2})R^{(2)}(\theta_1,\phi_1)I^{(3)}.
\end{equation}
To further simplify, we define
\begin{equation}
R^{(2)}(\theta_{mid},\phi_{mid}) = R^{(2)}(\theta_1,\phi_1+\frac{\pi}{2})R^{(2)}(\theta_1,\phi_1)
\end{equation}
so that
\begin{equation}
U = U_{4}U_{3}R^{(1)}(2\theta_1,\phi_1)R^{(2)}(\theta_{mid},\phi_{mid})I^{(3)}
\end{equation}
\begin{equation}
U = I^{(1)}R^{(1)}(2\theta_1,\phi_1)R^{(2)}(\theta_2,\phi_2)R^{(2)}(\theta_2,\phi_2)R^{(2)}(\theta_{mid},\phi_{mid})I^{(3)}I^{(3)}
\end{equation}
\begin{equation}
U = R^{(1)}(2\theta_1,\phi_1)R^{(2)}(2\theta_2,\phi_2)R^{(2)}(\theta_{mid},\phi_{mid})I^{(3)}.
\end{equation}
We see that with an appropriate choice for $\theta_1$, $\phi_1$, $\theta_2$, $\phi_2$ we can achieve any arbitrary rotations on ions 1 and 2 through this procedure with only four laser pulses. As a specific example, the choice $\phi_2 = \phi_{mid} + \pi$ and $\theta_2 = \theta_{mid}/2$ yields an arbitrary rotation on ion 1 without affecting ions 2 and 3:
\begin{equation}
U = R^{(1)}(2\theta_1,\phi_1)R^{(2)}(\theta_{mid}-2\theta_2,\phi_{mid})I^{(3)}
\end{equation}
\begin{equation}
U = R^{(1)}(2\theta_1,\phi_1)I^{(2)}I^{(3)}.
\end{equation}
Similar sequences extend this idea to even longer chains, although these require larger confinement changes to produce useful displacements between nearest-neighbor ions \cite{James1998} and correspondingly more laser pulses.

\section{Experimental Apparatus}\label{sec:Apparatus}
The ion-trapping apparatus previously used for the experiments described in \cite{deLaubenfels2015} was modernized to perform this work. It incorporates a surface-electrode ion trap with DC electrode potentials controlled by National Instruments 16-bit PXI-6733 digital-to-analog converter (DAC) cards clocked at 100 kHz. The output of each DAC is filtered with a 530 kHz reactive low-pass filter. The qubit for all of the experiments presented here is the $|S_{1/2},m_j=-^1/_2\rangle-|D_{5/2},m_j=-^1/_2\rangle$ transition in $^{40}$Ca$^+$. One-qubit rotations are achieved via optical pulses from an ultra-stable laser resonant with this transition at 729 nm. The ions are confined radially by a radio-frequency (RF) potential (peak magnitude $\approx 176$ V) at 56.4 MHz applied to the RF electrodes. A single ion in this potential is confined with an axial secular frequency $\omega_z = 2\pi\times 2.05$ MHz and radial frequencies $\omega_{r_1(r_2)} = 2\pi\times 7.89 (5.88)$ MHz. A bias magnetic field of $11.37$ G is provided by two rare-earth permanent magnets outside the vacuum chamber.

The GTRI-fabricated ``Gold Trap'' used here is an improved iteration of the GTRI Microwave Trap described in \cite{Shappert2013}; it is a planar linear Paul trap with 42 segmented electrodes, two outer rail electrodes, and integrated microwave waveguides. Ions are confined nominally 58 $\mu$m above the trap surface. In contrast with earlier versions, the segmented DC electrodes are located between the microwave waveguides to allow for stronger harmonic confinement. Gold vias (lateral dimension $20 \times 40$ $\mu$m) connect the top gold electrode layer to an underlying fan-out 1.5-$\mu$m thick aluminum layer. Each electrode benefits from 70 pF of capacitance to ground, an intentional side-effect of stray capacitance in the fanout layer, which reduces undesired RF pickup. The top layer of gold is designed with a moderately high aspect ratio to reduce variations in trapping potential caused by dielectric charging (e.g. from UV exposure). Specifically, the gold layer is 10 $\mu$m thick, and the nominal gap betweeen electrodes is 6 $\mu$m; it is fabricated via an electroplating process with a photoresist electroplating mold.

Both linear ion transport and axial potential modulation are effected with an assortment of waveforms applied to the DC electrodes. These waveforms are calculated using the methods described in \cite{Blakestad2011} with an additional quadrupole rotation designed to rotate the radial axis by $\approx 7.3^\circ$ to aid Doppler cooling of the vertical radial motional mode. The potential modulation required for a differential $\pi$ phase shift between the ions in a pair is achieved by scaling the DC potentials up or down, and we interpolate between these initial and final potential configurations with a user-defined number of intermediate configurations (waveform points). In practice we find that, in addition to uniformly scaling the potential, we must also apply a carefully tuned compensation electric field (most importantly along the crystal axis) in order to reduce motional excitation.

\section{Experimental Results}
\subsection{Individually Addressed Ramsey Experiments}\label{sec:Ramsey}
In order to demonstrate single-ion addressing via potential modulation (SIAPM), we first perform a pair of Ramsey experiments in which only one of two co-trapped ions is addressed. The ions are Doppler cooled initially via laser beams at 397 nm (red detuned from the $S_{1/2}-P_{1/2}$ transition) and 866 nm (blue detuned from the $D_{3/2}-P_{1/2}$ transition). The two axial modes of the ion pair are then sideband cooled to $\bar{n}\ll1$ via alternating laser beam pulses at 729 nm (resonant with the $S_{1/2}-D_{5/2}$ transition), 854 nm (resonant with the $D_{5/2}-P_{3/2}$ transition), and 866 nm (also resonant with the $D_{3/2}-P_{1/2}$ transition). A $\sigma^+$-polarized pulse at 397 nm followed by additional optical pumping with alternating 729 nm and 866/854 nm pulses initializes the population into a single $S_{1/2}$ Zeeman state with fidelity $\gg0.99$. 

The Ramsey sequences here consist of two composite $\pi/2$ gates separated by a delay. The composite $\pi/2$ gates are comprised of a $\pi/4$ optical pulse, a modulation in the trapping potential, and a second $\pi/4$ optical pulse as described in \sref{sec:Theory}. After the second $\pi/4$ optical pulse, we return the axial potential to its initial value (although this is not strictly necessary). Following the two composite Ramsey pulses, the ions are split into separate wells and illuminated alternately with final 397 nm pulses to detect their qubit states individually ($S_{1/2}$ fluoresces while $D_{5/2}$ remains dark).
 
\begin{figure}
	\centering
	\includegraphics[width=0.9\columnwidth]{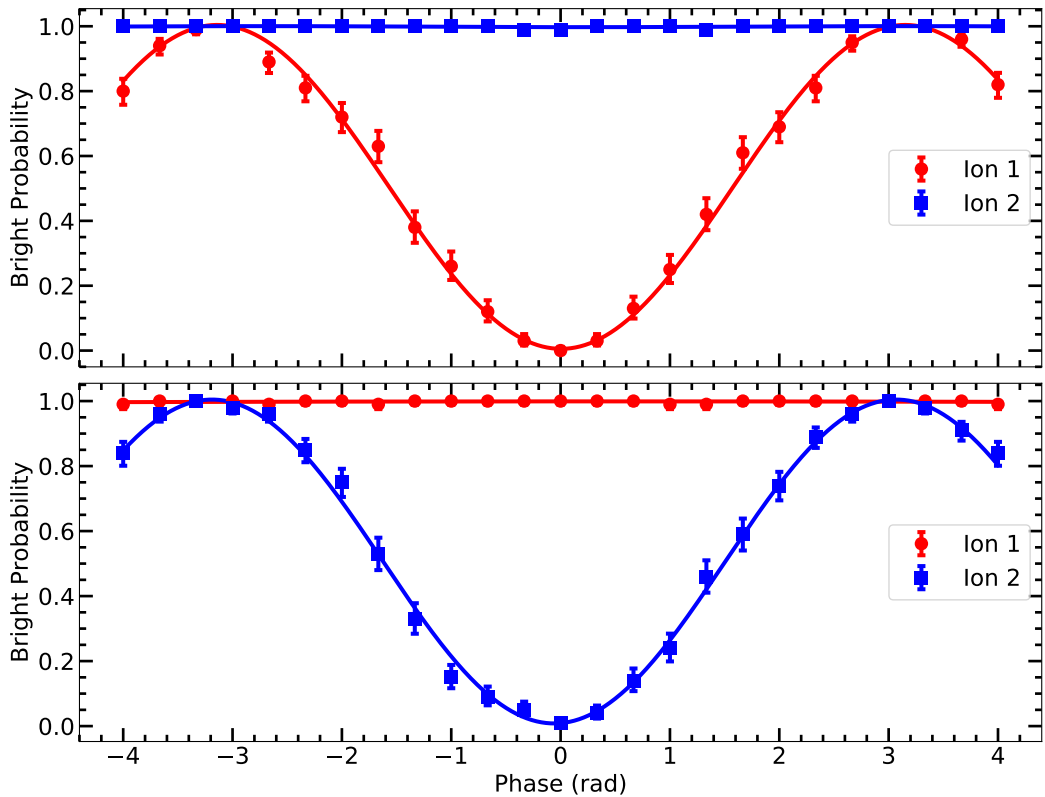} 
	\caption{SIAPM Ramsey experiment results. Red circles (blue squares) represent the data for ion 1 (2), and solid lines represent sinusoidal fits to the data. (Upper) Targeting ion 1 while performing the identity on ion 2. (Lower) Targeting ion 2 while performing the identity on ion 1. Error bars represent detection histogram statistical uncertainty.}
	\label{fig:IndAddRamseyInd}
\end{figure}

The results of these experiments are shown in \fref{fig:IndAddRamseyInd}. The upper panel presents the results of an experiment where ion 1 is intended to rotate and precess while ion 2 experiences no net rotation. The lower panel presents the results of the opposite situation. A least-squares fit of these data to a sinusoid yields a contrast of $0.999(10)$ when targeting ion 1 and $0.996(12)$ when targeting ion 2. In both datasets the non-targeted ion remains consistently bright (within the experimental errorbars) indicating that it experienced an identity operation.

\subsection{Individually Addressed Randomized Benchmarking}\label{sec:2IonRB}
To better quantify the errors induced by SIAPM in a computational context, we perform simultaneous one-qubit randomized benchmarking experiments on two ions within the same potential well. As in \sref{sec:Ramsey}, the two ions are first Doppler and (axially) sideband cooled, and initialized into a single $S_{1/2}$ Zeeman state. A random sequence of one-qubit operations (described below) is then applied, after which the ions are split into separate wells and measured.

The RB sequences are structured as described in \cite{Brown2011} and elaborated in \cite{Gaebler2012}. These sequences are composed of a given number (length) of steps, each of which consists in turn of a Pauli gate ($\pi$-rotations about the $X$ axis, $Y$ axis, $Z$ axis, or the identity) and a Clifford gate (comprised of between zero and three $\pi/2$-rotations about the $X$ axis or $Y$ axis with $1.5$ on average). The $X$-axis and $Y$-axis $\pi$ ($\pi/2$) gates are implemented as follows (see \sref{sec:Theory}): a $\pi/2$ ($\pi/4$) optical pulse, a modulation in the trapping potential, a second $\pi/2$ ($\pi/4$) optical pulse, and a return to the initial trapping potential. The $Z$-axis $\pi$ and identity gates are implemented simply by adjusting the phases of subsequent gates. For these experiments we interleave two independent, random sequences of steps, each sequence targeting only one of the two ions.

\begin{figure}
	\centering
	\includegraphics[width=0.75\columnwidth]{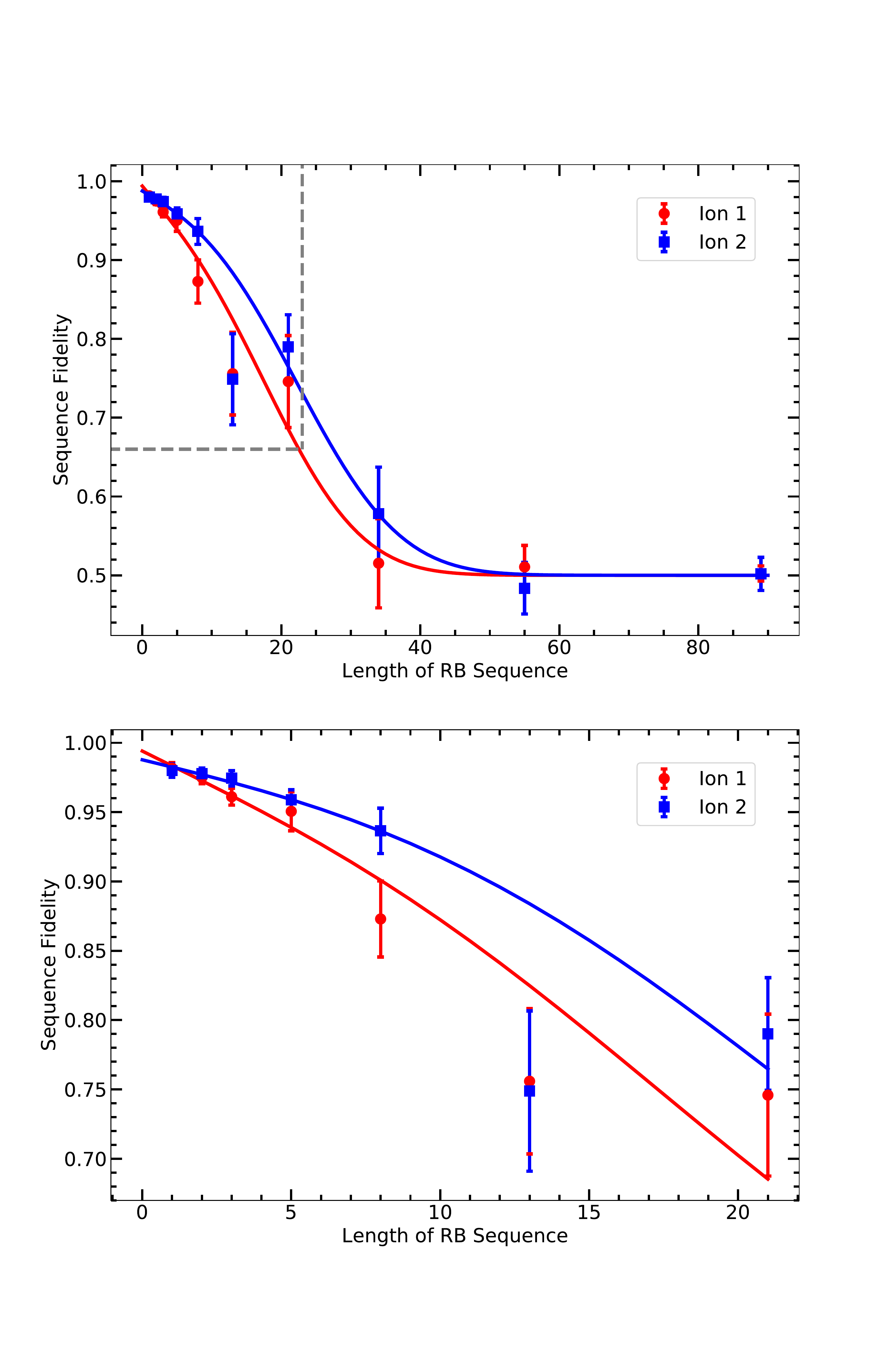} 
	\caption{Simultaneous, individually-addressed, one-qubit RB on two ions in the same potential well. Red circles (blue squares) represent the average fidelity for a given sequence length for ion 1 (2), and solid lines show fits of \eref{Eq:RBFit} to the data. (Upper) The full dataset. The grey dashed lines outline the area of a magnified view (lower) at shorter sequence lengths. Error bars represent the standard error of the mean.}
	\label{fig:2IonRB}
\end{figure}

The results of this experiment appear in \fref{fig:2IonRB}. The observed decay in average fidelity is not purely exponential; it falls off more quickly at long sequence lengths than at short ones, a behavior that we attribute to unwanted motional excitation (heating). On average, performing one RB step on each of the two ions requires eight optical pulses and eight potential modulations. Although these potential modulations are calibrated to minimize ion heating, some undesired excitation is present; we explore the origins of this heating in greater detail in \sref{sec:ErrorDiagnostics}. To account for this ion heating during longer RB sequences, we fit the data to an alternative function (derived in detail in \ref{app:FitFunction}) to that used in \cite{Brown2011,Gaebler2012}. A more rigorous statistical approach to fitting RB to alternative models such as this one is described in \cite{Proctor2019}. Assuming that the ion motional states are thermal, that the temperature grows linearly with RB sequence length, and that only the COM mode\footnote{The expression may be extended to include all motional modes. See \ref{app:FitFunction}.} is excited, we are led to a modified fidelity fit function
\begin{equation}
\mathcal{F} = \frac{1}{2} + \frac{1}{2}\left(1-2\varepsilon_{\rm SPAM}\right)\left(1-2\varepsilon_{\rm step}\right)^{l}\prod_{m = 1}^{l}\left[\left(1-2\varepsilon_{m,{\rm P}}\right)\left(1-2\varepsilon_{m,{\rm C}}\right)\right]
\label{Eq:RBFit}
\end{equation}
where $\varepsilon_{\rm SPAM}$ is the state preparation and measurement (SPAM) error, $\varepsilon_{\rm step}$ is the error per sequence step in the absence of heating, and $l$ is the sequence length. $\varepsilon_{m,{\rm P}}$ ($\varepsilon_{m,{\rm C}}$) is the error of the Pauli (Clifford) gate due to finite temperature at the $m^{\rm th}$ sequence step.

\begin{table}
	\caption{\label{tab:2IonRB}Individually-addressed RB fit coefficients. Data in \fref{fig:2IonRB} were fit to \eref{Eq:RBFit}. Fit-coefficient errors represent statistical fit uncertainty.}
	\begin{indented}
		\item[]\begin{tabular}{@{}llll}
			\br
			& $\varepsilon_{\rm SPAM}$ & $\varepsilon_{\rm step}$ & $\Delta\bar{n}$\\
			\mr
			Ion 1 & $6(6)\times10^{-3}$		& $1.1(3)\times10^{-2}$	& $4.0(1.2)$\\
			Ion 2 & $1.2(5)\times10^{-2}$ 	& $5(2)\times10^{-3}$	& $3.4(7)$	\\
			\br
		\end{tabular}
	\end{indented}
\end{table}

We have fit the data in \fref{fig:2IonRB} to \eref{Eq:RBFit} with the results listed in \tref{tab:2IonRB}. The expected spontaneous emission $(\Gamma_{D5/2}\sim1.2 {~\rm s}^{-1})$ contribution to the SPAM error is $6.7\times10^{-3}$ ($1.0\times10^{-2}$) for ion 1 (2), consistent with the experimental results. The per-step error is comparable to what we observe typically in experiments with a single ion (\sref{sec:ErrorDiagnostics}) since each step in the individually addressed case contains four times the number of optical pulses. The remaining discrepancy we attribute to miscalibrations in ion position when modulating the potential. Each step in the RB sequence includes eight potential modulation cycles on average. From the fit results to our model including motional excitation, we estimate $\approx0.5$ quanta of COM mode excitation per modulation cycle.

\subsection{Error Diagnostics}\label{sec:ErrorDiagnostics}
\begin{figure}
	\centering
	\includegraphics[width=0.75\columnwidth]{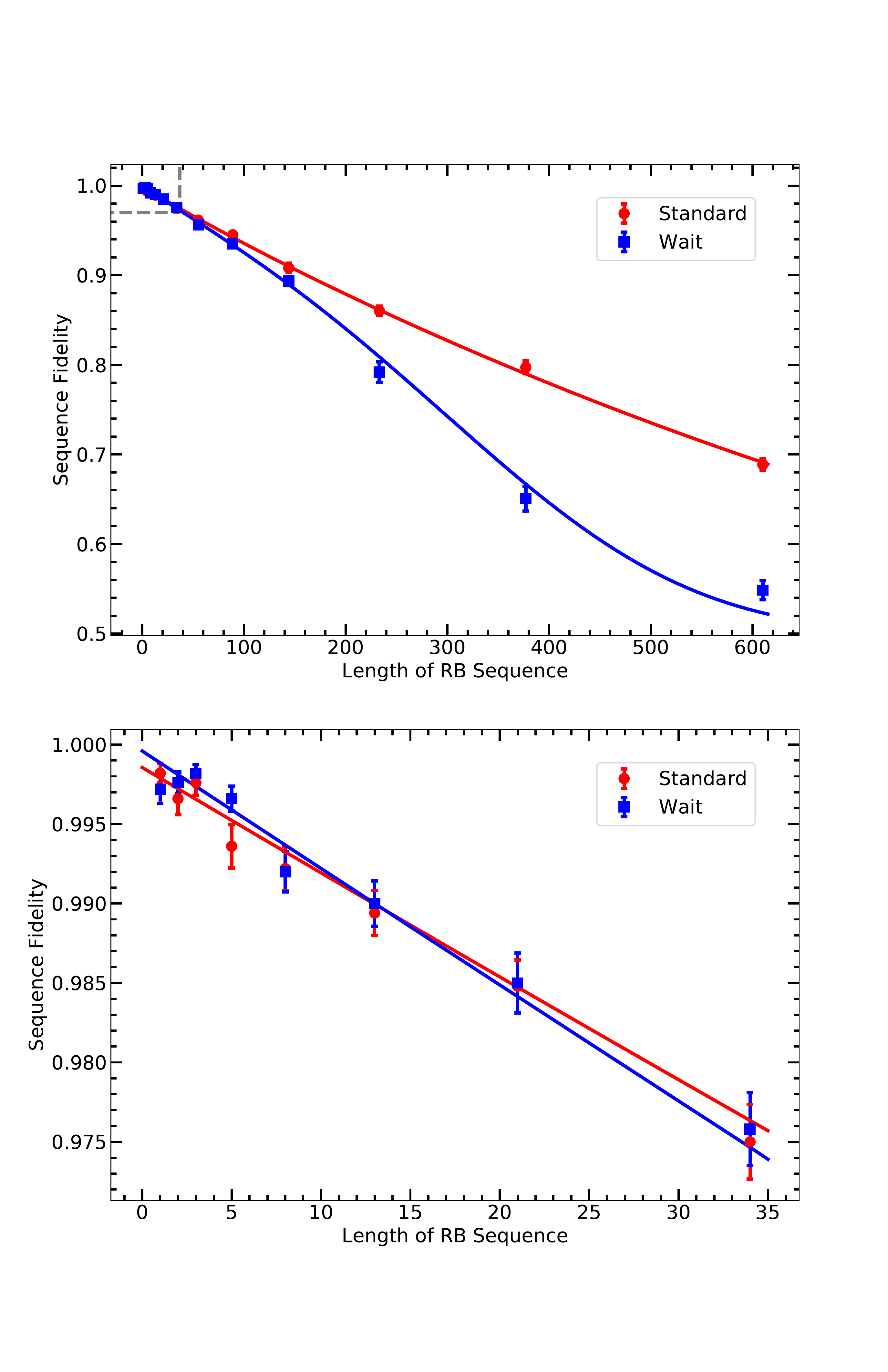} 
	\caption{Single-ion RB results. Red circles (blue squares) represent the average fidelity obtained without (with) inserting an additional 25 $\mu$s delay between gate pulses, and solid lines show fits of \eref{Eq:RBFit} to the data. (Upper) The full dataset. The grey dashed lines outline the area of a magnified view (lower) at shorter sequence lengths. Error bars represent the standard error of the mean.}
	\label{fig:CompRB2}
\end{figure}

To better characterize the errors observed in the two-ion, individually addressed RB results, we perform RB experiments using only a single ion both without and with the addition of a time delay after each optical gate for comparison. For both of these experiments, we use two $\pi/2$ pulses for the Pauli $\pi$ rotations and construct the Clifford gates from (non-composite) $\pi/2$ pulses without any modulation in trap potential. Other experimental details are identical to those in the two-ion experiments. The red data (``Standard'') in \fref{fig:CompRB2} gives the results of the experiment with no additional time delay after each optical gate.

In comparison with the two-ion results, the one-ion data agree far better with a simple exponential model. We nevertheless fit the data as before, with the results listed in \tref{tab:CompRB} (``Standard'' row). The per-step error of $6.7(2)\times10^{-4}$ is below the two-ion per-step error by more than an order of magnitude and forms a good baseline for the earlier results: we should expect the one-ion rate to be a lower-bound for the two-ion rate. We believe that residual laser noise and magnetic-field noise are the limiting factors in our one-ion fidelity. The observed SPAM error is consistent with spontaneous emission limits. The fit motional excitation from background heating is $6.4(1.2)\times10^{-3}$ quanta/step, which corresponds to an axial heating rate of $0.26(5)$ quanta/ms when taking into account the average duration for each step in a sequence. This value is roughly consistent with independent measures of the axial heating rate ($0.16(5)$ quanta/ms) made via the more conventional technique of observing red and blue sideband ratios \cite{Turchette2000}.

\begin{table}
	\caption{\label{tab:CompRB}Single-ion RB fit coefficients. Data in \fref{fig:CompRB2} were fit to \eref{Eq:RBFit}. Fit-coefficient errors represent statistical fit uncertainty.}
	\begin{indented}
		\item[]\begin{tabular}{@{}lllll}
			\br
			& & & & Fitted Heating\\
			RB Variation & $\varepsilon_{\rm SPAM}$ & $\varepsilon_{\rm step}$ & $\Delta\bar{n}$ & Rate (q/ms)\\
			\mr
			Standard	& $1.4(4)\times10^{-3}$	& $6.7(2)\times10^{-4}$ & $6.4(1.2)\times10^{-3}$	& $0.26(5)$\\
			Wait		& $4(4)\times10^{-4}$	& $7.4(3)\times10^{-4}$ & $2.7(2)\times10^{-2}$		& $0.29(2)$\\
			\br
		\end{tabular}
	\end{indented}
\end{table}

The two-ion individually addressed RB sequences include various modulations in the trapping potential which significantly increase the overall sequence durations beyond those of the one-ion case. These additional delays could be expected to contribute significant errors. For our second one-ion diagnostic experiment we intentionally insert an additional 25 $\mu$s delay after each optical pulse. This delay is chosen to match the duration of the potential modulations in the two-ion experiment (``Wait'' RB data in \fref{fig:CompRB2}; fit coefficients in \tref{tab:CompRB}).

We observe only a small increase in gate error from $6.7(2)\times10^{-4}$ to $7.4(3)\times10^{-4}$ due to the additional delay. Again we can estimate an axial heating rate and obtain $0.29(2)$ quanta/ms, in agreement with the previous value and still roughly consistent with independent measures. This consistency strengthens our confidence in the modified randomized benchmarking fit function \eref{Eq:RBFit} and indicates that these fits represent a previously unexplored method to quantify ion-trap heating rates. The higher heating rate observed in RB experiments could be due to the contributions of radial-mode heating, which would not affect the axial sideband ratio measurement.

\section{Outlook and Conclusion}
We have individually addressed each ion in a two-ion string using axial potential modulation and a global gate beam, and we have characterized the resulting gate performance via simultaneous randomized benchmarking using independent sequences for each ion. We have also compared our SIAPM results to the results of a series of one-ion experiments in order to better understand our primary sources of gate error. This experiment is a prerequisite for the universal control of two ions confined in the same potential well. The SIAPM technique we used avoids several of the challenges associated with other individual addressing methods and is broadly applicable to CCD-style ion trap architectures \cite{Kielpinski2002} where only two or three ions are co-trapped at a time. Our technique also has utility as a diagnostic in QPT or RB experiments used to evaluate the performance of multi-qubit entangling gates, where it can be implemented at lower cost than can alternatives involving tightly focused beams.

While these demonstrations do not represent state-of-the-art one-qubit gate fidelities, the SIAPM technique works with sufficient fidelity at small numbers of gates to perform useful quantum process tomography of a two-qubit gate. To extend this technique to longer gate sequences (needed e.g. for two-qubit gate randomized benchmarking), we will suppress the heating induced by the potential modulations. Here we anticipate improvements through a combination of (1) more careful calibration, (2) finer waveform control achieved via faster DACs, and (3) precompensation of the waveforms for the low-pass filter response. Theoretically, SIAPM fidelity is limited only by the number and fidelity of the laser pulses involved: the simplest two-pulse sequence implementing a single gate on a two-ion chain should incur an error only twice as high as that of the individual pulses. Higher pulse fidelities in our apparatus can be achieved through improvements to the qubit laser linewidth and via improved suppression of background magnetic field variations.

Similarly, the modest speed of the sequences used in these demonstrations is currently constrained by the ${\approx}100$ kHz sample rate of our DACs, but this is not a fundamental limit. Adiabatic modulations of the trapping potential can be achieved in a few microseconds with faster DACs, while even faster diabatic changes can in principle be realized with finely sampled waveforms \cite{Bowler2012,Walther2012,Alonso2016}. Because the laser pulses typically require durations of several microseconds, it should be possible to perform SIAPM sequences during intervals only about twice as long as that of an individual pulse.

\ack
The authors thank D. Leibfried for stimulating discussions. Research was sponsored by the Army Research Office and was accomplished under Grant Number W911NF-18-1-0166. The views and conclusions contained in this document are those of the authors and should not be interpreted as representing the official policies, either expressed or implied, of the Army Research Office or the U.S. Government. The U.S. Government is authorized to reproduce and distribute reprints for Government purposes notwithstanding any copyright notation herein. Trap fabrication was performed in part at the Georgia Tech Institute for Electronics and Nanotechnology, a member of the National Nanotechnology Coordinated Infrastructure (NNCI), which is supported by the National Science Foundation (ECCS-1542174).

\appendix
\section{RB Fit Function}\label{app:FitFunction}
\subsection{Fidelity At Finite Mode Temperature}
We define the fidelity at the end of an RB sequence as
\begin{equation}
\mathcal{F}={\rm Tr}\left(\rho_{\rm actual}\rho_{\rm ideal}\right)
\end{equation}
where $\rho_{\rm actual}$ ($\rho_{\rm ideal}$) is the actual (intended) density matrix after an operation. For a one-qubit rotation of some angle $\theta_0$, we have
\begin{equation}
\rho_{\rm ideal}=\left[\matrix{\frac{1}{2}\left(1+\cos\left(\theta_0\right)\right)&\frac{1}{2}\sin\left(\theta_0\right)\cr\frac{1}{2}\sin\left(\theta_0\right)&\frac{1}{2}\left(1-\cos\left(\theta_0\right)\right)\cr}\right].
\end{equation}
If the ion does not occupy the ground state of motion, but instead occupies a Fock state $n$ of a motional mode, then
\begin{equation}
\rho_{\rm actual}=\left[\matrix{\frac{1}{2}\left(1+\cos\left(\theta_n\right)\right)&\frac{1}{2}\sin\left(\theta_n\right)\cr\frac{1}{2}\sin\left(\theta_n\right)&\frac{1}{2}\left(1-\cos\left(\theta_n\right)\right)\cr}\right].
\end{equation}
where $\theta_n=\theta_0L_n\left(\eta^2\right)$ is the reduced rotation angle due to the finite Fock state \cite{Leibfried2003}, $L_n$ is the Laguerre polynomial of order $n$, and $\eta$ is the Lamb-Dicke parameter for the given motional mode. The fidelity is then 
\begin{equation}
F_n=\frac{1}{2}\left[1+\cos\left(\theta_0-\theta_n\right)\right]=\frac{1}{2}\left[1+\cos\left(\theta_0\left(1-L_n\left(\eta^2\right)\right)\right)\right].
\end{equation}
Thermally averaging the above expression with the Boltzmann weighting function
\begin{equation}\label{eqn:thermal}
W_n=\frac{1}{1+\bar{n}}\exp\left[-\ln\left(1+\frac{1}{\bar{n}}\right)n\right],
\end{equation}
results in the expression for the one-qubit rotation fidelity:
\begin{equation}\label{eq:chiintro}
\mathcal{F}=\sum_{n=0}^{\infty}W_nF_n=\frac{1}{2}\left[1+\chi\left(\theta_0,\bar{n},\eta\right)\right]
\end{equation}
where
\begin{equation}\label{eq:chidef}
\chi\left(\theta_0,\bar{n},\eta\right)=\sum_{n=0}^{\infty}W_n\cos\left(\theta_0\left(1-L_n\left(\eta^2\right)\right)\right).
\end{equation}

The above expressions can be generalized to a series of $l$ one-qubit rotations with an increasing motional temperature. The fidelity expression then becomes
\begin{equation}
\mathcal{F}=\frac{1}{2}\left[1+\prod_{m = 1}^{l}\chi\left(\theta_0,\bar{n}_m,\eta\right)\right]
\end{equation}

\subsection{$\eta^2\ll1$ Limit}
We know of no analytic solution to a thermal average of Laguerre polynomials as in \eref{eq:chiintro} and \eref{eq:chidef} (these expressions can still be used for numerical fits); making the assumption that $\eta^2\ll1$ ($L_n\left(\eta^2\right)\approx1-n\eta^2$), the Boltzmann-averaged fidelity can be expressed as
\begin{equation}
\mathcal{F}\left(\theta_0,\bar{n},\eta\right)=\frac{1}{2}\left[1+\frac{1}{2\left(\bar{n}+1\right)-\left[\frac{2\bar{n}+1}{1+\bar{n}\left(1-\cos\left(\theta_0\eta^2\right)\right)}\right]}\right].
\end{equation}

\subsection{Multiple Motional Modes}
To account for multiple motional modes of $N$ ions,
\begin{equation}
\theta_n\rightarrow\theta_0\prod_{k=1}^{3N}L_{n_k}\left(\eta^2_k\right).
\end{equation}
The fidelity of a one-qubit rotation with a given set of Fock state occupations is given by
\begin{equation}
F_{n_1,n_2,...,n_{3N}}=\frac{1}{2}\left[1+\cos\left(\theta_0\left(1-\prod_{k=1}^{3N}L_{n_k}\left(\eta^2_k\right)\right)\right)\right].
\end{equation}
Here, the Boltzmann weighting function includes all motional modes:
\begin{equation}
W_{n_1,n_2,...,n_{3N}}=\prod_{k=1}^{3N}\frac{1}{1+\bar{n}_k}\exp\left[-\ln\left(1+\frac{1}{\bar{n}_k}\right)n_k\right].
\end{equation}
The resulting expression for the one-qubit rotation fidelity is then
\begin{equation}
\mathcal{F}=\sum_{n_1}^{\infty}\sum_{n_2}^{\infty}...\sum_{n_{3N}}^{\infty}W_{n_1,n_2,...,n_{3N}}F_{n_1,n_2,...,n_{3N}}=\frac{1}{2}\left[1+\chi\left(\theta_0,\boldsymbol{\bar{n}},\boldsymbol{\eta}\right)\right]
\end{equation}
where
\begin{eqnarray}
\fl \chi\left(\theta_0,\boldsymbol{\bar{n}},\boldsymbol{\eta}\right)=\sum_{n_1}^{\infty}\sum_{n_2}^{\infty}...\sum_{n_{3N}}^{\infty}W_{n_1,n_2,...,n_{3N}}\cos\left(\theta_0\left(1-\prod_{k=1}^{3N}L_{n_k}\left(\eta^2_k\right)\right)\right)
\end{eqnarray}
and bold type denotes variables with $3N$ components.

Using the $\eta^2\ll1$ limit described above, the Boltzmann-averaged fidelity can be expressed as
\begin{eqnarray}
\fl F_{n_1,n_2,...,n_{3N}}\approx\frac{1}{2}\left[1+\cos\left(\theta_0\left[1-\left(1-n_1\eta_1^2\right)\left(1-n_2\eta_2^2\right)...\left(1-n_{3N}\eta_{3N}^2\right)\right]\right)\right].
\end{eqnarray}
Keeping only leading-order terms in $\eta_k$,
\begin{equation}
F_{n_1,n_2,...,n_{3N}}\approx\frac{1}{2}\left[1+\cos\left(\theta_0\left[n_1\eta_1^2+n_2\eta_2^2+...+n_{3N}\eta_{3N}^2\right]\right)\right].
\end{equation}
The notation of the above expression can be simplified by converting to and taking the real part of complex exponentials (we note that a similar mathematical treatment is presented in the appendix of \cite{Roos2000}):
\begin{equation}
F_{n_1,n_2,...,n_{3N}}\approx\frac{1}{2}\left[1+\mathcal{R}\left\{\prod_{k=1}^{3N}\exp\left(i\theta_0n_k\eta_k^2\right)\right\}\right].
\end{equation}
The thermally-averaged fidelity can then be simplified to
\begin{equation}
\mathcal{F}\approx\frac{1}{2}\left[1+\mathcal{R}\left\{\prod_{k=1}^{3N}\left(\frac{1}{\bar{n}_k+1}\right)\frac{1}{1-\frac{\bar{n}_k}{\bar{n}_k+1}\exp\left(i\eta_k^2\theta_0\right)}\right\}\right].
\end{equation}
Here,
\begin{equation}
\chi\left(\theta_0,\boldsymbol{\bar{n}},\boldsymbol{\eta}\right)\approx\mathcal{R}\left\{\prod_{k=1}^{3N}\left(\frac{1}{\bar{n}_k+1}\right)\frac{1}{1-\frac{\bar{n}_k}{\bar{n}_k+1}\exp\left(i\eta_k^2\theta_0\right)}\right\}.
\end{equation}

\subsection{RB Fit Function}
To account for the ion heating during longer RB sequences, we use a function adapted from that used in \cite{Brown2011,Gaebler2012}. Assuming that the ion motional states are thermal \eref{eqn:thermal}, that the temperature grows linearly with RB sequence length, and that only the COM mode is excited, we are led to a modified fidelity fit function
\begin{equation}
\mathcal{F} = \frac{1}{2} + \frac{1}{2}\left(1-2\varepsilon_{\rm SPAM}\right)\left(1-2\varepsilon_{\rm step}\right)^{l}\prod_{m = 1}^{l}\left[\left(1-2\varepsilon_{m,{\rm P}}\right)\left(1-2\varepsilon_{m,{\rm C}}\right)\right]
\end{equation}
where $\varepsilon_{\rm SPAM}$ is the state preparation and measurement (SPAM) error, $\varepsilon_{\rm step}$ is the error per sequence step in the absence of heating, and $l$ is the sequence length. $\varepsilon_{m,{\rm P}}$ ($\varepsilon_{m,{\rm C}}$) is the error of the Pauli (Clifford) gate due to finite temperature at the $m^{\rm th}$ sequence step. We model the temperature as $\bar{n}_m = \bar{n}_0 + m\cdot\Delta\bar{n}$, which describes the initial temperature $\bar{n}_0$ (we measure $\bar{n}_0\sim0.01$) and heating of $m\cdot\Delta\bar{n}$ at step $m$. 

Each step in the RB sequence contains on average one $\pi$ or identity gate and one and a half $\frac{\pi}{2}$ gates. However, not all gate implementations incur motional excitation errors. Since only two of the four Pauli gates incur errors due to motional excitation and the gates are chosen at random with equal probability, we choose
\begin{equation}
\left(1-2\varepsilon_{m,{\rm P}}\right)=\frac{1}{2}\left[1+\chi\left(\pi,\bar{n}_m,\eta\right)\right].
\end{equation}
Therefore, the error per Pauli gate on average is half what we would have calculated if we had assumed that each Pauli gate incurred the same error.

For the Clifford gate, we pick at random with equal probability from the identity, $X\left(\frac{\pi}{2}\right)$, $Y\left(\frac{\pi}{2}\right)$, $X\left(\frac{\pi}{2}\right)Y\left(\frac{\pi}{2}\right)$, $Y\left(\frac{\pi}{2}\right)X\left(\frac{\pi}{2}\right)$, and $X\left(\frac{\pi}{2}\right)Y\left(\frac{\pi}{2}\right)X\left(\frac{\pi}{2}\right)$. Correspondingly, we choose
\begin{eqnarray}
\fl \left(1-2\varepsilon_{m,{\rm C}}\right)=\frac{1}{6}\left[1+2\chi\left(\frac{\pi}{2},\bar{n}_m,\eta\right)+2\chi\left(\frac{\pi}{2},\bar{n}_m,\eta\right)^2+\chi\left(\frac{\pi}{2},\bar{n}_m,\eta\right)^3\right].
\end{eqnarray}

\section*{References}
\bibliography{SIAPM_Paper}
\bibliographystyle{iopart-num}
\end{document}